\documentclass[twocolumn,superscriptaddress,10pt]{revtex4-1}

\usepackage[dvips]{graphicx}
\usepackage{latexsym}
\usepackage{amsmath}
\usepackage{amssymb}
\usepackage{amsfonts}
\usepackage{color}
\usepackage{bm}
\usepackage{verbatim}
\usepackage[english]{babel}
\usepackage[utf8]{inputenc}
\usepackage{euscript}
\usepackage{amsbsy}
\usepackage{subfigure}
\usepackage{tikz}
\usepackage{tikz}

\graphicspath{{../pics/}{.}}

\renewcommand{\div}{\mathop{\mathrm{div}}}

\newcommand{\curl}{\mathop{\mathrm{curl}}}

\begin{document}

%
\title{
    Laser pulse probe of the chirality of Cooper pairs}

\author{V. L. Vadimov}
\author{A. S. Mel'nikov}
\affiliation{Institute for Physics of Microstructures, Russian
Academy of Sciences, 603950 Nizhny Novgorod, GSP-105, Russia}
\affiliation{University of Nizhny Novgorod, 23 Gagarin Avenue, 603950 Nizhny Novgorod, Russia}

\begin{abstract}
The internal chirality of Cooper pairs is shown to modify strongly the response of a superconductor to the local heating by a laser beam.
The suppression of the chiral order parameter inside the hot spot appears to induce the supercurrents flowing around the spot region.
The chirality affects also the sequential stage
of thermal quench developing according to the Kibble-Zurek scenario: besides the generation of
 vortex--antivortex pairs  the quench facilitates the formation of superconducting domains with different chirality.
These fingerprints of the chiral superconducting state  can be probed by any experimental
techniques sensitive to the local magnetic field.
The supercurrents encircling the hot spot originate from the inhomogeneity of the state with the broken time reversal symmetry and their detection
would provide a convenient alternative to the search of the spontaneous edge currents sensitive to the boundary properties.
 Thus, the suggested setup can help to resolve
 the long-standing problem of unambiguous detection of type of pairing in $\mathrm{Sr_2RuO_4}$ considered as a good candidate
 for chiral superconductivity.
\end{abstract}

\maketitle

\section{Introduction}
The interaction of light with different types of orderings in condensed matter
systems is in the focus of current research in the field of
optoelectronics~\cite{refKirilyuk,refKimel,refFausti2011light,refSuda743,refVeshchunov2016optical}.
The  light controlled manipulation of magnetic and/or superconducting states
provides a perspective way to construct new fast operating switching
devices~\cite{refKirilyuk,refKimel} and
serves a basis for different experimental methods probing and characterizing
the order parameter structure and dynamics~\cite{refAltshuler,
refMatsunaga1145}.
In particular, remarkable progress has been recently achieved in the design of sensitive superconducting bolometers
and photon detectors \cite{refSemenov2001349,refGoltsman}. Fundamental issues of the order parameter dynamics have been investigated probing the Higgs mode
in the superconducting state \cite{refAltshuler,refMatsunaga1145}.

The simplest physical picture describing the effect of the light pulse on superconductor can be constructed starting
 from a so called ``hot spot'' model~\cite{refSemenov2001349,refMaingault,refZotovaPRB}. Within this approach one can assume the energy of the laser pulse to be transferred to the electronic subsystem
which results in further formation of the locally heated state with an increased electronic temperature. This increase in the local temperature
is responsible for the partial or complete suppression of the superconducting
order parameter in the region of the hot spot. The state with the
inhomogeneous temperature is unstable due to the heat diffusion and at the
next stage the hot spot disappears and  the order parameter relaxes to its
initial value before the light pulse absorption. The exact
picture depends of course on the electron-electron and electron-phonon relaxation rates
which are responsible for different stages of the evolution of the
nonequilibrium electronic distribution. Provided the relaxation stage is
rather short and can be considered as a rapid thermal quench the thermodynamic
order parameter fluctuations can complicate the return to the initial state
giving rise to the formation of the vortex - antivortex pairs according to the
Kibble-Zurek scenario~\cite{refKibble,refZurek,refVolovikKZ, refManivPolturak}.

Should we expect any essential changes in the above model if the superconducting order parameter
is not just a single
complex function but may have several components or possess a nontrivial anisotropy in the momentum space? In other words, does the study of the superconductor dynamics excited by the light pulse allows to distinguish the states with different internal structure of the Cooper pairs?
The goal of the present paper is to develop a theoretical basis for the use of the laser beam as a probe of such unconventional superconducting states,
more specifically the states with a nonzero internal average angular momentum
of the Cooper pairs $\mathbf{L}$\footnote{The projections of the angular momentum are bad
quantum numbers in the crystals due to the lack of continuous rotational
symmetry. The true structure of the gap is described by the basis functions of
some irreducible representation of the crystal symmetry group which
corresponds to the certain superconducting state. Still these
functions can be classified by the average value of the angular momentum.}. It is instructive to start our discussion
 of the appropriate generalization of the hot spot model from a qualitative
 analysis of inhomogeneous states for $\mathbf{L}\neq 0$.   The angular
 momentum of the relative motion of two electrons in the pair naturally
 interacts with the angular momentum of the motion of the pair center of mass.
 For any inhomogeneity of the superconducting state this interaction of the
 angular momenta can induce the supercurrents and corresponding magnetic
 field.
 Naively, one can expect these supercurrents to be proportional to the
 effective magnetization currents of the Cooper pairs: $\propto \curl \mathbf{L}$.
 However the orbital angular momentum of the sample bulk appears to be
 significantly
 reduced compared to expected $N\hbar / 2$ value where $N$ is the total amount
 of electrons in a volume. The orbital momentum of the
 bulk is determined by the contribution of the Cooper pairs which are smaller
 than the inter--pair distance so the momentum is reduced by factor $(T_c/
 E_F)^2$~\cite{refVolovik1981orbital}. A more precise analysis gives additional
 logarithmic factor and final expression has a form $L_z \propto \hbar N (T_c
 / E_F)^2 \ln (E_F /
 T_c)$~\cite{refVolovik1975angular,refCross1975generalized}.
 A noticeable contribution 
to the supercurrent is provided by another mechanism, namely by the mixture of several order parameter components generated by the order parameter inhomogeneity. Such mixture originates from the obvious fact that the projection of the internal angular momentum can not conserve in the presence of the
inhomogeneity. Previously, the search of the corresponding spontaneous
currents was mainly related to the edge of the samples. Unfortunately, near
the edge the order parameter inhomogeneity and, thus, the generation of the
additional order parameter components and the resulting supercurrents 
strongly depend on the details of the electron scattering at the surface, i.e.
on the surface quality~\cite{refSauls,refLederer,refHuang2,refBakurskiy}. As a consequence, the edge currents can be strongly
diminished and surface imperfections may cause the difficulties in their
experimental observation in $\mathrm{Sr_2RuO_4}$~\cite{refKirtley} which is considered to be
a most promising candidate for a superconductor with the chiral $p$-wave order
parameter~\cite{refKallin}. {
    The other explanations of the edge currents absence focus on possibility
    of chiral non $p$-wave pairing type in
    $\mathrm{Sr_2RuO_4}$~\cite{refScaffidi} for which the macroscopic orbital
    momentum vanishes in the finite size
    samples~\cite{refHuang,refTada,refVolovikChiral}. Also the properties of
    the edge currents appear to depend on the band structure of 
    $\mathrm{Sr_2RuO_4}$~\cite{refBouhon,refImai}.
}

\begin{figure}
    \includegraphics{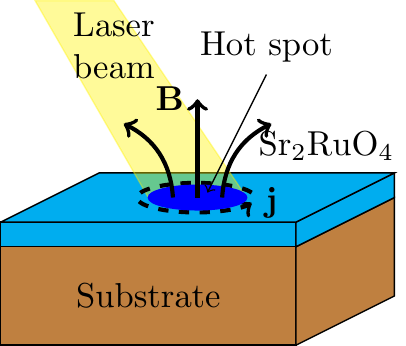}
    \label{figSetup}
    \caption{(Color online) The proposed experimental setup. The laser beam
    heats the sample inducing the supercurrents flowing around the hot spot.
    The magnetic field created by these currents can be observed by the SQUID
    microscope or the Hall sensor.}
\end{figure}

From this point of view the order parameter inhomogeneity created by the laser
beam far from the sample edge of uncontrolled quality provides much better
conditions for the study of the above supercurrents. The current circulating
around the hot spot (see Fig.~\ref{figSetup}) should be easily detected by any
experimental techniques sensitive to small local magnetic field such as SQUID
microscope or sensitive Hall sensor. 
The generation of the magnetic field in the
inhomogeneously heated samples is common for the systems with the broken time
reversal symmetry. For example the magnetic field appears in the hot spots in
multiband $s+is$ and $s+id$ superconductors~\cite{refSilaevBTRS}.
The generation of the secondary order parameter components can be even more
pronounced at the late stage of the hot spot evolution. Indeed, provided the
spot dimension exceeds the coherence length one can expect that the rather
fast quench to the initial temperature can be accompanied by nucleation of
different order parameter components forming, thus, a domain structure in
addition to the well-known vortex-antivortex configurations induced by the
Kibble--Zurek mechanism.

For the further quantitative analysis of the above phenomena we choose a
rather general two-component Ginzburg--Landau model introduced previously in a
number of works~\cite{refHeeb,BarashMelnikovZhETFEn} for the description of
properties of  the $\mathrm{Sr_2RuO_4}$ compound~\cite{refHeeb,VadimovSilaevPhysRevLett111p177001}.

In the Section II we introduce Ginzburg--Landau free energy and the main
equations describing the superconducting state. For the case of a smooth order
parameter profile in the hot spot these equations are solved in the Section
III. The final expression for the magnetic field is
obtained for the Gaussian profile of temperature. The Section IV is devoted to
 superconductor dynamics in the thermal quench mode and study of
the chiral domain generation according to the Kibble--Zurek mechanism. The
stability of the domains and their interaction with the vortices is discussed
in the Section V. In the final Section VI we sum up the results of this paper.

\section{Model}

The superconducting order parameter has two components $(\Psi_+, \Psi_-)$
which stand for the Cooper pairs with opposite direction of internal orbital
momentum. The free energy of the superconductor is given by the following
expression\cite{refHeeb}:
\begin{multline}
    F =  \int \left\{-a \left(|\Psi_+|^2 + |\Psi_-|^2\right) + \frac{b_1}{2}\left(|\Psi_+|^2 +
    |\Psi_-|^2\right)^2 + \right. \\ b_2 \left|\Psi_+ \Psi_-\right|^2 + \\
    K_1 \left(|D_+ \Psi_+|^2 + |D_- \Psi_+|^2 + |D_+ \Psi_-|^2 + |D_-
    \Psi_-|^2\right) + \\
    K_2 | D_+ \Psi_- + D_- \Psi_+|^2 + \\ K_3 \left [ |D_+ \Psi_+|^2 + |D_-
    \Psi_-|^2 + (D_+ \Psi_-)^\ast (D_- \Psi_+) + c.c. \right] +\\
    \left.\frac{\left(\curl \mathbf A\right)^2}{8\pi}\right\}\;d^3 r\ ,
\end{multline}
where $\mathbf A$ is
the vector potential of the magnetic field, $\mathbf D = -i \nabla - 2 \pi /
\Phi_0 \mathbf A$ is the covariant derivative, $D_\pm = (D_x \pm i D_y) / \sqrt{2}$, $\Phi_0$ is the superconducting
flux quantum, $a$, $b_1$, $b_2$, $K_1$, $K_2$ and $K_3$ are the
phenomenological parameters. The coefficient $a$ depends on the temperature as
follows $a = \alpha (T_c - T)$. For simplicity we omitted the terms lowering
the symmetry of the free energy to $D_{4h}$
symmetry, restricting ourselves to the case of $D_{6h}$ crystal.Also we
assume that the spatial variations of the order parameter are only in $xy$
plane and neglect the variations along $z$-axis. If $b_2 > 0$ the equilibrium homogeneous states have form of chiral
domains $(\Psi_+,
\Psi_-) = (\sqrt{a / b_1}, 0)$ and $(\Psi_+, \Psi_-) = (0, \sqrt{a / b_1})$.

The laser pulse is absorbed by the electron subsystem of the superconductor
inducing, thus, non-equilibrium distribution of the quasiparticles in the
sample. In general case this distribution does not correspond to any
temperature though if we suppose that the electron--electron scattering
is the fastest process in the system the distribution of the quasiparticles
rapidly thermalizes. Then one can introduce an inhomogeneous temperature 
$T(\mathbf r)$ and the parameter $a$ also becomes a function of the
coordinates $a = \alpha(T_c - T(\mathbf r))$. We suppose inhomogeneity to have a form of
a hot spot and the temperature to be 
constant far from it, i.e. $a(\mathbf r) \to a_0 = \alpha (T_c - T_0)$ at $r
\to \infty$, where $T_0$ is the bath temperature.
Introducing the dimensionless order parameter components $\eta_\pm = \Psi_\pm
\sqrt{b_1 / a_0}$ we rewrite the free energy:
\begin{multline}
    \label{eq_gl_dimensionless}
    F = \frac{H_{cm}^2}{4\pi}\int \left\{-\tau(|\eta_+|^2 + |\eta_-|^2) + \frac{1}{2}
    \left(|\eta_+|^2 + |\eta_-|^2\right) + \right. \\ \beta |\eta_+ \eta_-|^2 +
    \xi^2 \left(|\mathbf D \eta_+|^2 + |\mathbf D \eta_-|^2\right) + \\ 2 \xi^2
    \zeta \left[ (D_+ \eta_-)^\ast (D_- \eta_+) + c.c.\right] +\\ \left.
        \frac{\left(\curl \mathbf A\right)^2}{2 H_{cm}^2} \right\}\;d^3 r\ ,
\end{multline}
where $H_{cm}= \sqrt{4\pi a_0^2 / b_1}$ is the thermodynamical critical field,
$\xi = \sqrt{(K_1 + K_2) b_1} / a_0$ is the coherence length, $\tau(\mathbf r)
= a(\mathbf r) / a_0$ and $\zeta = K_2 /
(K_1 + K_2)$ and $\beta = b_2 a_0^2 / b_1^2$. Here we assume that $K_2 -
K_3$ is negligible due to the small electron--hole
asymmetry\cite{refBalatksiMineev}.

One can obtain the Ginzburg--Landau equations for the order parameter
components:
\begin{multline}
    \xi^2\left(\mathbf D^2 \eta_+ + 2 \zeta D_+^2 \eta_-\right) - \tau \eta_+ +
    \eta_+ |\eta_+|^2 + \\ (1 + \beta) | \eta_-|^2 \eta_+ = 0 \ ,
\end{multline}
\begin{multline}
    \xi^2\left(\mathbf D^2 \eta_- + 2 \zeta D_-^2 \eta_+\right) - \tau \eta_- +
    \eta_- |\eta_-|^2 + \\ (1 + \beta) | \eta_+|^2 \eta_- = 0 \ .
\end{multline}
Varying the free energy functional over the vector potential $\mathbf A$ one
can obtain the expression for the
superconducting current:
\begin{multline}
    \label{eq_current}
    \mathbf j_s = -\frac{c \Phi_0}{16 \pi^2 \lambda^2} \left\{
        \eta_+^\ast (\mathbf D \eta_+)
        +
        \eta_-^\ast (\mathbf D \eta_-) + \phantom{\sqrt{2}}\right. \\
        \left.
        \zeta \sqrt{2} (\hat{\mathbf x}+ i \hat{\mathbf y}) \left[\eta_- (D_-
        \eta_+)^\ast + \eta_+^\ast (D_+ \eta_-)\right] + c.c.
        \right\} \ ,
\end{multline}
where $\lambda = \Phi_0  / [4 \sqrt{2\pi^3 (K_1 + K_2)}]$ is the London penetration length.
The first terms proportional to $\eta_+^\ast(\mathbf D \eta_+)$ and
$\eta_-^\ast(\mathbf D \eta_-)$ are common
for the Ginzburg--Landau theory of conventional superconductors.
The rest two terms  contain not only the contributions proportional to the
superfluid velocities of the different order parameter components but also a
non-zero contribution caused by the inhomogeneity of the magnitudes of the
order parameter components.
Below we show that the suppression of one of the order
parameter components can generate another component and the corresponding
superconducting current.

\section{Weak heating. Adiabatic approximation}

We choose for definiteness a chiral domain with $\eta_+ = 1$ and $\eta_- = 0$
and consider a hot spot located far from the domain boundaries.
To elucidate our main results we start from a simplified ``adiabatic'' model
assuming 
that the temperature varies slowly at the length scale $\xi$, i.e.
$|\nabla \tau| \ll \tau / \xi$.
Under these assumptions
the dominating order parameter component $\eta_+$ ``follows'' the local temperature and 
the other order parameter component $\eta_-$ can be found as a perturbation:
\begin{gather}
    \eta_+ \approx \sqrt{\tau} e^{i\phi} \ ,\\
    \eta_- \approx -\frac{2 \zeta \xi^2 D_-^2 \eta_+}{\beta \tau} \ ,
\end{gather}
where $\phi$ is the unknown phase. Also we assume that the
sample is a thin film with the thinkness $d$ much smaller than the London
penetration length $\lambda$. This simplification allows us to neglect the vector
potential and the screening effects. For simplicity we assume
that the temperature distribution is
axi-symmetric $\tau(\mathbf r) = \tau(r)$. Then we can omit the phase $\phi$
and find the order parameter components in the following form:
\begin{gather}
    \eta_+(r) = \sqrt{\tau} = f_+(r)\\
    \eta_-(r, \varphi) = \frac{\zeta \xi^2}{\beta \tau}
    \left(\frac{\partial^2 \eta_+}{\partial r^2} - \frac{1}{r} \frac{\partial
    \eta_+}{\partial r}\right) e^{-2 i\varphi}  = f_-(r) e^{-2i \varphi}\ ,
\end{gather}
where $r$ and $\varphi$ are the polar coordinates. Now we can substitute the
order parameter components into the expression~(\ref{eq_current}) obtain the following
expression for superconducting current, neglecting the terms of order
$\eta_-^\ast
\nabla \eta_-$:
\begin{equation}
    j_\varphi \approx -\frac{c \Phi_0 \zeta}{8 \pi^2 \lambda^2} \left(f_+
    \frac{\partial f_-}{\partial r} - f_- \frac{\partial f_+}{\partial r} +
    \frac{2}{r} f_+ f_-\right) \ .
\end{equation}
The current has only the azimuthal component. This current creates the magnetic
field which can be measured experimentally. The value of the field in the
center of the spot can be found as follows:
\begin{equation}
    \label{eq_field_adiabatic}
    B_z = -\frac{\Phi_0 \zeta}{4\pi
    \lambda_{\mathit{eff}}}\int\limits_0^\infty\frac{dr}{r} \left(f_+
    \frac{\partial f_-}{\partial r} - f_- \frac{\partial f_+}{\partial r} +
    \frac{2}{r} f_+ f_-\right) \ ,
\end{equation}
where $\lambda_{\mathit{eff}} = \lambda^2 / d$ is the effective penetration
length. The magnetic field far from the defect has a dipole form with the magnetic
moment equal to:
\begin{equation}
    m = -\frac{\Phi_0 \zeta}{8\pi
    \lambda_{\mathit{eff}}}\int\limits_0^\infty r^2 \left(f_+
    \frac{\partial f_-}{\partial r} - f_- \frac{\partial f_+}{\partial r} +
    \frac{2}{r} f_+ f_-\right) \; dr\ .
\end{equation}

The above approach, indeed, can be applied only if the temperature varies slowly
and is always below the critical one. Otherwise one should solve the
Ginzburg--Landau equations numerically.

\subsection{Gaussian beam}

The evolution of the local temperature is a complicated process which is governed by the heat diffusion, electron--phonon
interaction and nonequilibrium phonons escape to the
substrate~\cite{refGulyan1985electron}.
The diffusion can be neglected if all other characteristic times like time of
the electron--phonon interaction and phonon escape time are much less than the
characteristic diffusion time which depends on the beam size. With the
diffusion omitted the local temperature appears to be a function of
the local absorbed power which can be linearized in the vicinity of the bath
temperature $T_0$ for the weak heating.

Assuming a Gaussian profile of the laser beam we have the
following temperature profile:
\begin{equation}
    T(\mathbf r) = T_0 + \frac{\kappa P}{2\pi \sigma^2}
    \exp\left(-\frac{r^2}{2 \sigma^2}\right) \ ,
\end{equation}
where $\kappa$ is the proportionality coefficient between the local power and
the temperature. One can introduce the dimensionless power $\tau_0 = \kappa P
/ [\pi \xi^2 (T_c - T_0)]$ and obtain the following expression for the order
parameter components, magnetic field in the spot and the magnetic moment for a slightly
heated spot $(T - T_0) \ll (T_c - T_0)$:
\begin{gather}
    \eta_+ \approx 1 - \frac{\tau_0 \xi^2}{2 \sigma^2} e^{-r^2 / 2\sigma^2}
    - \frac{\tau_0^2 \xi^4}{8 \sigma^4}
    e^{-r^2 / \sigma^2}\ , \\
    \eta_- \approx -\frac{\zeta}{\beta}\left[\frac{\tau_0 \xi^4
    r^2}{2\sigma^6} e^{-r^2/2\sigma^2} + \frac{\tau_0^2 \xi^6 r^2}{\sigma^8}
e^{-r^2/\sigma^2}\right]\\
    B_z \approx \frac{\Phi_0}{4 \pi \lambda_{\mathit{eff}} \xi} \cdot
    \frac{7 \zeta^2 \tau_0 \sqrt{\pi}}{8 \beta}
    \left(\frac{\xi}{\sigma}\right)^5 \ , \\
    m \approx \frac{\Phi_0 \xi^2}{8\pi \lambda_{\mathit{eff}}} \cdot
    \frac{\zeta^2 \tau_0^2}{2 \beta} \left(\frac{\xi}{\sigma}\right)^4 \ .
\end{gather}
The magnetic
field and the magnetic moment reveal power-law dependence on the beam size
$B_z \propto
\sigma^{-5}$, $m \propto \sigma^{-4}$ and intensity $B_z \propto P$, $m
\propto P^2$.
\begin{figure*}[t]
    \centering
    \subfigure[]{
        \includegraphics[width=0.47\textwidth]{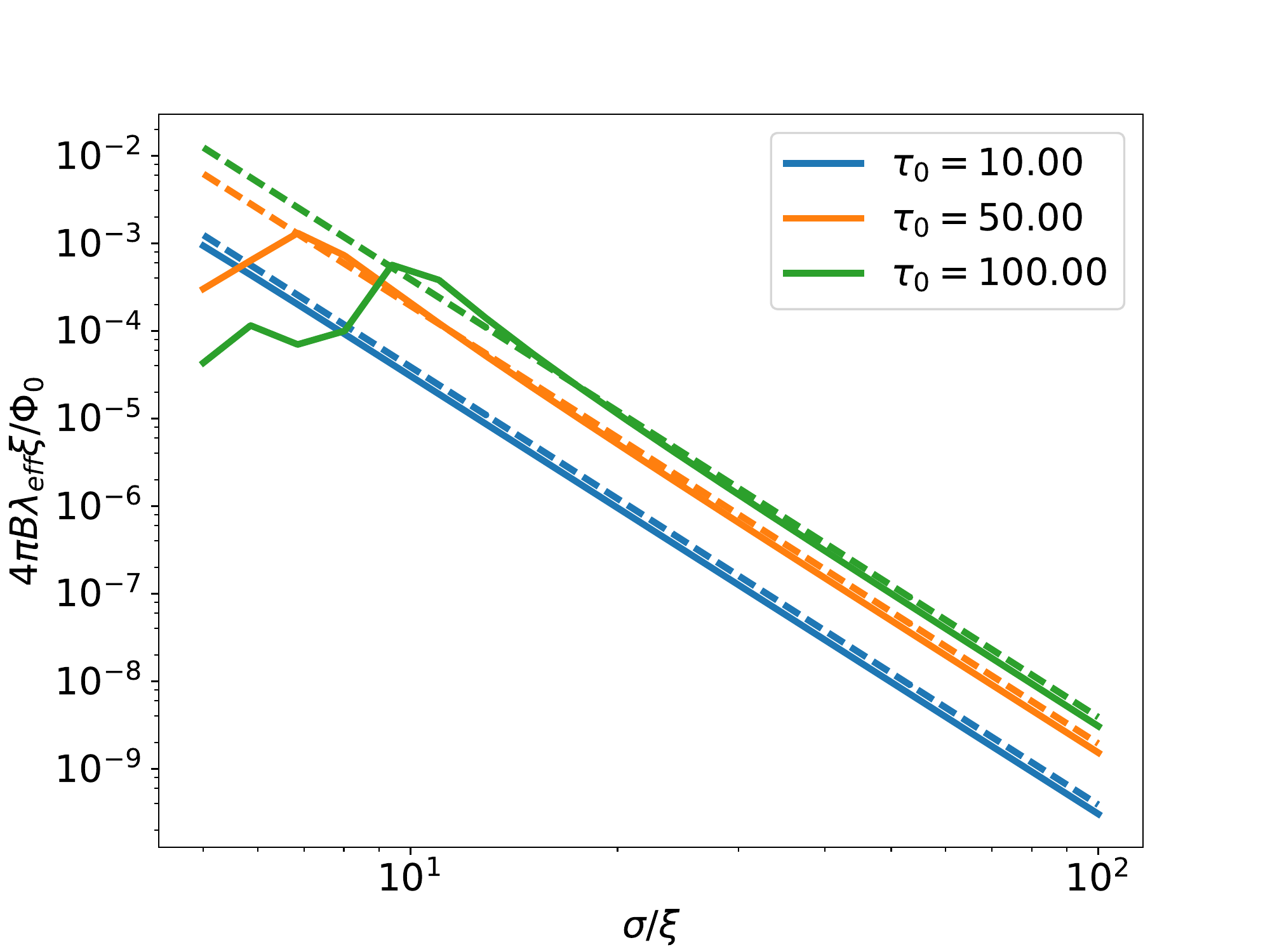}
    }
    \subfigure[]{
        \includegraphics[width=0.47\textwidth]{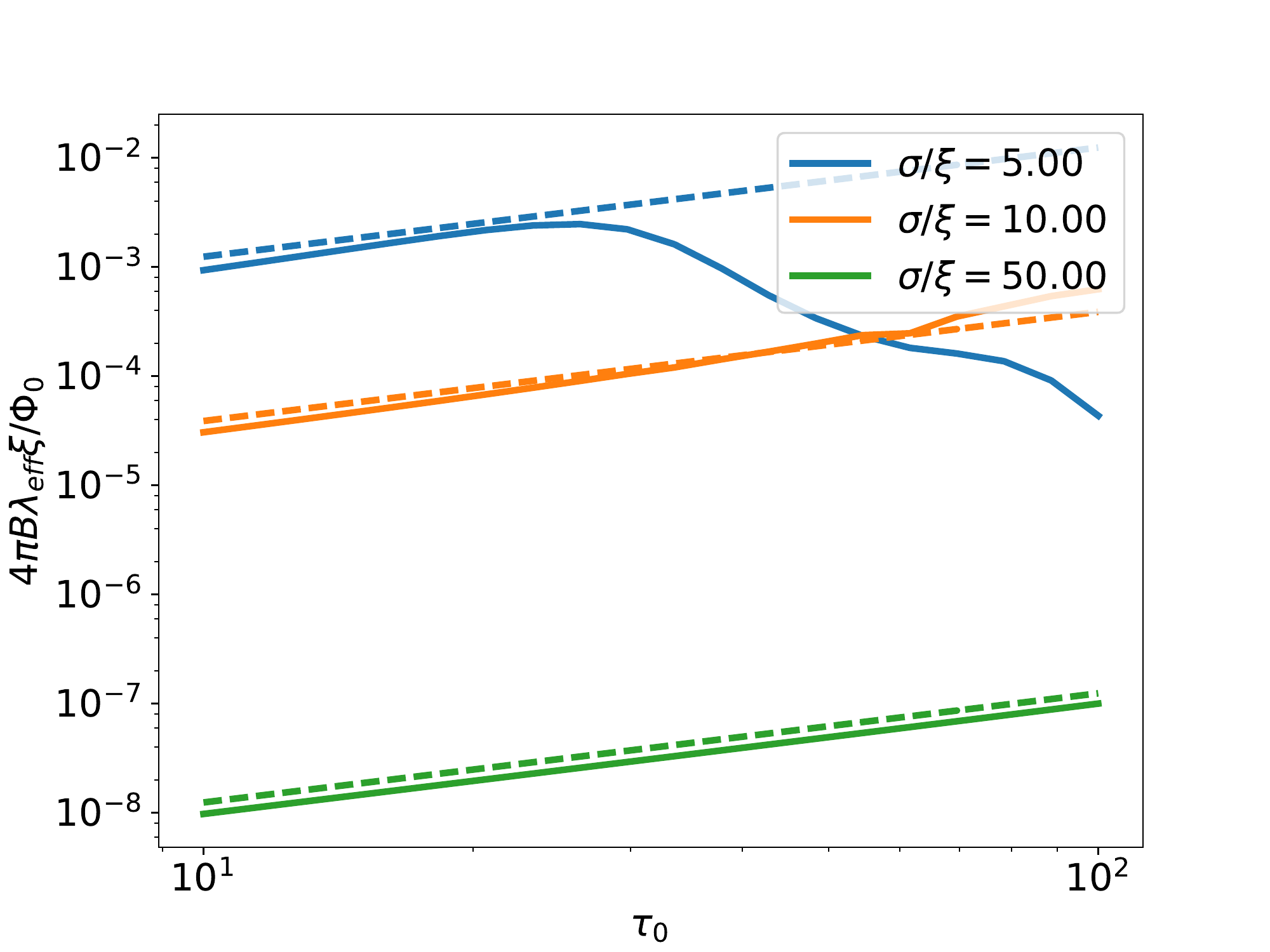}
    }
    \subfigure[]{
        \includegraphics[width=0.47\textwidth]{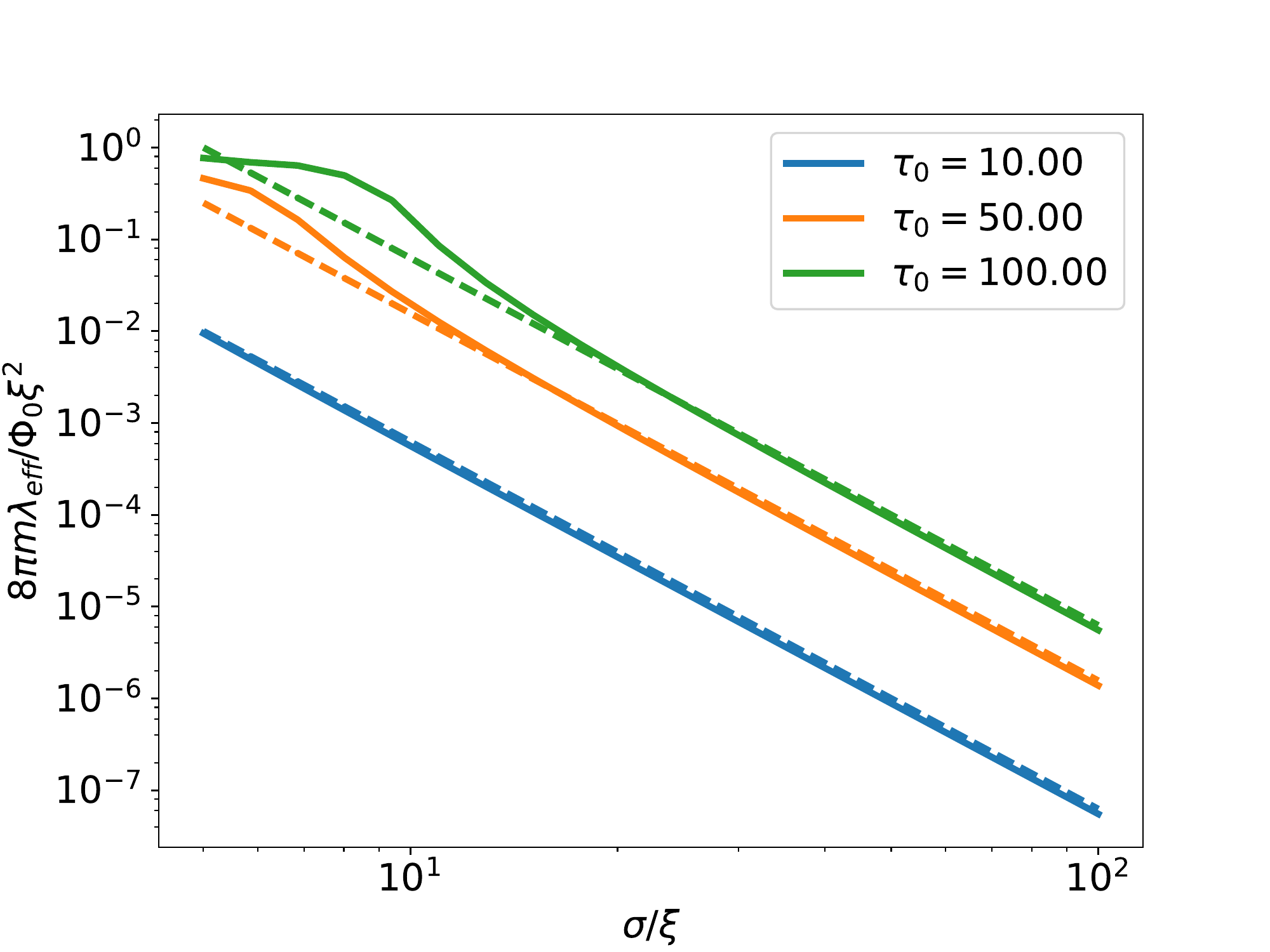}
    }
    \subfigure[]{
        \includegraphics[width=0.47\textwidth]{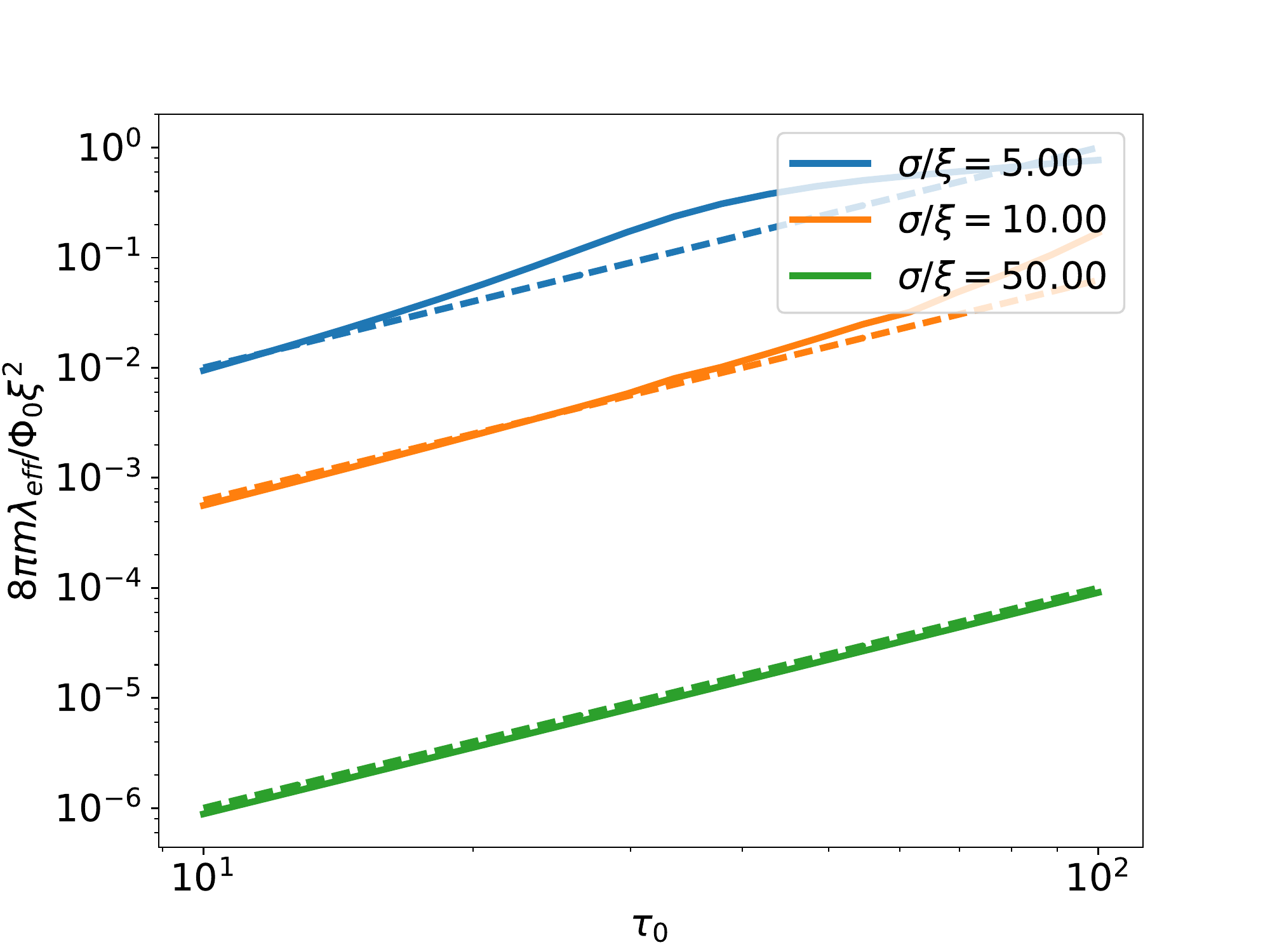}
    }

    \caption{(Color online) The magnetic field in the center of the spot (a), (b) and the magnetic
    moment (c), (d) vs the beam size (a), (c) and intensity (b), (d). The
    solid line correspond to the solutions obtained numerically. The
    results of the adiabatic approximation for the weak field are represented
    by the dashed lines. The parameters of the Ginzburg--Landau functional are
    $\beta = 1$ and $\zeta = 0.5$, no screening is considered.}
    \label{figRes}
\end{figure*}

The magnetic field in the center of the spot and the
magnetic moment of the currents versus the beam parameters $\sigma$ and $\tau_0$
are shown on the Fig.~\ref{figRes}. The adiabatic approximation works
reasonably for the slow temperature variations and fails if the local
temperature is close to the critical one.

The dependence of the magnetic field in the
center of the spot on the beam size $\sigma$ and the dimensionless power
$\tau_0$ calculated for the Gaussian spot within the adiabatic approach~(\ref{eq_field_adiabatic})
and numerically. The adiabatic approximation is reasonable for the slow
temperature variations and fails if the local temperature is close to the
critical temperature.

The maximal value of the field is reached when temperature in the center of
the spot is close to the critical one. The field is given in the units of $H_0
= \Phi_0 / (4\pi \lambda_{\mathit{eff}} \xi) = H_{cm} / \sqrt{2} \cdot d /
\lambda$ on the Fig.~\ref{figRes}(a) and (b). The plot shows that the maximal
field achieved for the small spots is between $10^{-2} \cdot H_0$ and $10^{-3}
\cdot H_0$. 
At the border of applicability range we suppose that the thickness of the
film is $d = \lambda$ and consider low temperature parameters $\xi =
66$~nm and $\lambda = 152$~nm. These assumptions give us an estimate of
observable field up to $1.5$~G. In fact this value may be too optimistic due
to the Meissner screening which comes into play for the thick enough samples.

{The similar generation of the magnetic field in the hot spots may also occur in the other
superconductors with the broken time reversal symmetry like $s+id$
superconductors~\cite{refSilaevBTRS}. However the patterns of the magnetic
field in
the $s+id$ and the chiral $p$-wave superconductors
 appear to be qualitatively different due to the different symmetry of the
 superconducting states. Assuming an axially symmetric
 temperature distribution we find that the supercurrent and, thus, the magnetic
 field also has the axial symmetry. We neglected the terms in the
 free energy which reduce the symmetry of the superconductor to $D_{4h}$ so
 the pattern of the magnetic field is expected to have tetragonal distortions. This
 still qualitatively differs from the case of $s+id$ superconductor in which the
 pattern of the magnetic field has a pronounced two-fold
 symmetry~\cite{refSilaevBTRS}. Thus, these types of pairing may be
 distinguished using the spatially resolved magnetic field measurements.}

 {Another fingerprint of the chiral superconductivity is a nonzero magnetic
moment of the thermally induced currents in the superconducting films. 
The key difference between the $s+id$ and $p_x+ip_y$ states is that the latter
is characterized by an internal vorticity in the momentum space directed along
the $z$ axis. Thus, in contrast to the $s+id$-wave superconductors the total
magnetic moment of the induced currents (integrated over the sample) can be
nonzero for a hot spot in a p-wave superconductor and the magnetic moment
direction should depend on the internal vorticity
 which is proved by the above direct
calculations. So magnetic moment measurements provide another possibility of
provide of the chiral $p$-wave superconductivity.}

\section{Strong heating. Domain generation}
In the case of the strong heating the temperature within the spot may exceed the
critical temperature of the superconductor. This results in the significant
suppression of the order parameter components. In this case one cannot apply the above
perturbation approach directly 
however the qualitative picture is similar: there is a supercurrent flowing
around the normal spot which creates magnetic field.  
However for a pulsed heat up the
relaxation of temperature can cause the formation of the chiral domains 
 according to the Kibble--Zurek mechanism
\cite{refKibble,refZurek,VadimovSilaevPhysRevLett111p177001}. The domain walls
carry superconducting current\cite{refMatsumoto} which can be detected by
the techniques sensitive to the magnetic field. The same mechanism is responsible for creation of
vortex--antivortex pairs in non-equilibrium transitions in $s$-wave
superconductors~\cite{refManivPolturak} which can be identified by the specific magnetic field
pattern. In the case of chiral $p$-wave superconductors the pattern
of the magnetic field appears to reveal a number of specific features which
can be used to distinguish this type of pairing.

We are using the approximation of the local temperature assuming now that it
can depend on time. We start from the strongly non-homogeneous temperature
distribution
which gradually relaxes to the equilibrium value $T = T_0$. We studied the
growth of the chiral domains numerically within the time-dependent Ginzburg--Landau approach: 
\begin{multline}
    -t_{\mathit{GL}} \left(\frac{\partial}{\partial t} + \frac{2\pi i c}{\Phi_0}
    \varphi\right) \eta_+ = \chi_+ (\mathbf r, t) - \tau(\mathbf r, t) \eta_+
    + \\ \eta_+
    |\eta_+|^2 + \eta_+ |\eta_-|^2 ( 1 + \beta) + \xi^2\left( \mathbf D^2 \eta_+ +
    2 \zeta D_+^2 \eta_-\right) \ ,\\
\end{multline}
\begin{multline}
    -t_{\mathit{GL}} \left(\frac{\partial}{\partial t} + \frac{2\pi i c}{\Phi_0}
    \varphi\right) \eta_- = \chi_- (\mathbf r, t) - \tau(\mathbf r, t) \eta_-
    + \\ \eta_-
    |\eta_-|^2 + \eta_- |\eta_+|^2 ( 1 + \beta) + \xi^2\left( \mathbf D^2 \eta_- +
    2 \zeta D_-^2 \eta_+\right) \ ,\\
\end{multline}
\begin{gather}
    \sigma_n \nabla^2 \varphi + {c} \div \frac{\delta F}{\delta
    \mathbf A} = 0 \ ,\\
    \frac{\sigma_n}{c} \frac{\partial \mathbf A}{\partial t} + \frac{c}{4\pi}\curl
    \curl \mathbf A + \sigma_n \nabla \varphi + c \frac{\delta
    F}{\delta \mathbf A} = 0 \ .
\end{gather}
The Coulomb gauge $\div \mathbf A = 0$ is considered, $\sigma_n$ is the
normal state conductivity, $t_{GL} = \Gamma / a_0$ is the order parameter
relaxation time and $\Gamma$ is a temperature independent constant.
The functions $\chi_\pm$ are
the delta--correlated noise sources $\langle \chi_\alpha(\mathbf r,  t)
\chi_\beta(\mathbf r',  t')\rangle = \chi^2 \delta_{\alpha_\beta} \delta(\mathbf r -
\mathbf r') \delta(t - t')$. Here we assume the thickness of the
superconducting film to exceed the penetration length $\lambda$ but to be
small
enough so that the sample could be heated homogeneously in $z$ direction. These simplifications allow us to consider 2D Meissner screening instead of solving the
full 3D problem.
The heat equation was not taken
into account selfconsistently. Instead the explicit model spatial and temporal profile of
temperature was specified $\tau = 1 - \frac{\tau_0 \xi^2}{\sigma^2}
\exp\left( -r^2 / [2 \sigma^2] - t / t_{\mathit{T}}\right)$, where
 $t_{\mathit{T}}$ is the
temperature relaxation time which is determined, e.g., by the heat flow into the
substrate.

If the temperature quench is adiabatically slow ($t_{T} \ggg t_{GL}$)
then the order parameter adiabatically follows the quasiequilibrium solution which is
slightly disturbed by the thermal fluctuations. In this case the homogeneous
domain appears after the quench is over.
However if the temperature quench has the similar rate as the 
order parameter relaxation rate ($t_{T} \sim t_{GL}$) then the state of the
superconductor is essentially non-equilibrium till the late stage of the
quench. The nuclei of both order parameter components arise from the thermal
fluctuations and grow rapidly until they are stabilized by the nonlinear terms in
the Ginzburg--Landau equation. The order parameter relaxation time $t_{GL}$
diverges at the temperatures close to the critical one $t_{GL} \propto (T_c -
T_0)^{-1}$ so the domain nucleation is likely to occur in the vicinity of the
phase transition.

\begin{figure*}
    \subfigure[]{
        \includegraphics[height=0.48\textwidth]{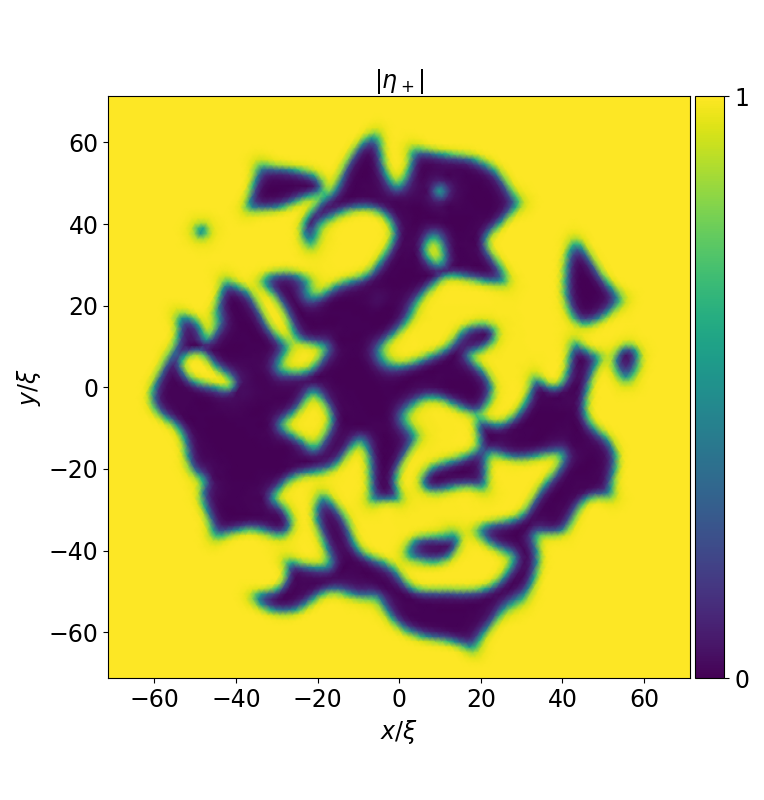}
    }
    \subfigure[]{
        \includegraphics[height=0.48\textwidth]{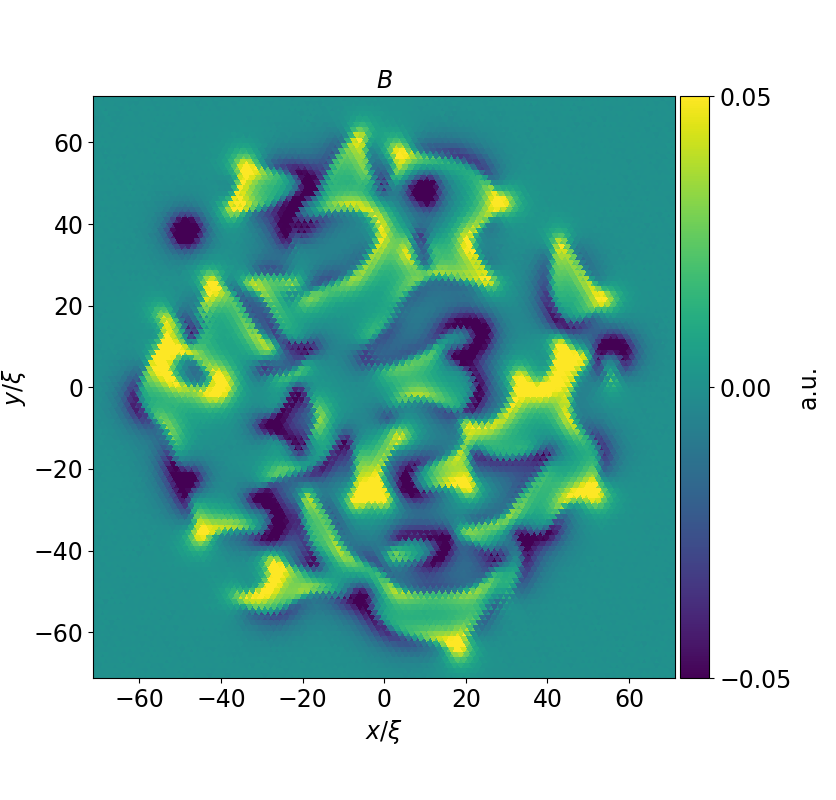}
    }
    \subfigure[]{
        \includegraphics[height=0.48\textwidth]{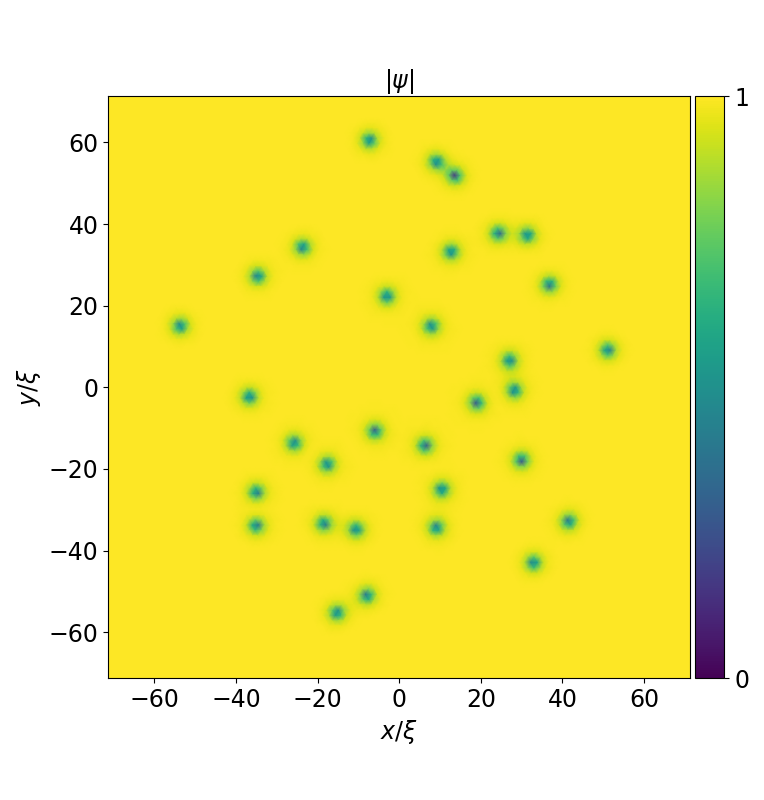}
    }
    \subfigure[]{
        \includegraphics[height=0.48\textwidth]{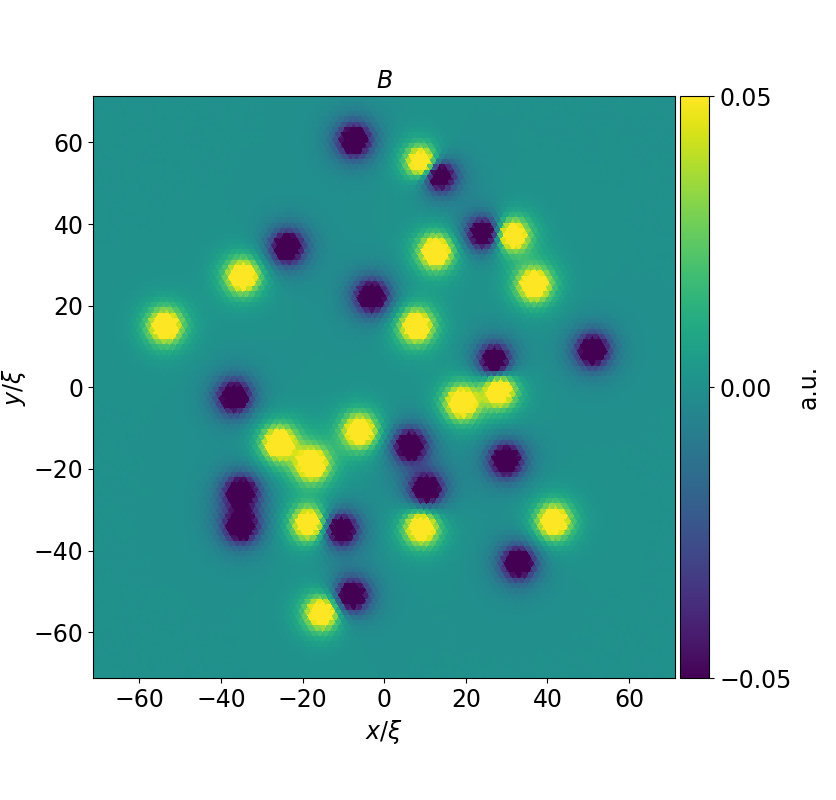}
    }
    \caption{(Color online) The results of the Kibble--Zurek quench for
    a $p$-wave superconductor (a,b) and a $s$-wave superconductor described by a simple
    one--component Ginzburg--Landau model (c,d). (a,c) The absolute values of
    the superconducting order parameters. 
    On the panel 
    (a) only the component $\eta_+$ is shown (the component $\eta_-$ is
    dominant in the areas where $\eta_+$ is
    suppressed). (b,d) The pattern of the magnetic field within the sample
after the quench. The magnetic field outside the film is expected to be
partially
smoothed out.}
    \label{fig_domains}
\end{figure*}

The results of the simulation are shown in the Fig.~\ref{fig_domains} (a,b). The
peculiar picture of the chiral domains appears after a
long time of simulation when the temperature is stabilized. The currents of
the domain
structure generate the inhomogeneous
magnetic field pattern with the zero total flux. The distribution of the magnetic
field qualitatively differs from the case of conventional $s$-wave
superconductor
for which Kibble--Zurek mechanism is known to be responsible for generation of
vortex--antivortex pairs\cite{refManivPolturak} (see Fig.~\ref{fig_domains}
(c,d)). { One can expect that the generation of the domain structure should be
accompanied by the generation of the vortex--antivortex pairs in the bulk of the domains but most of the
vortices appear to be pinned at the domain walls. The pinned vortices can be
found in the Fig.~\ref{fig_domains}(b) as asymmetric peaks of magnetic field. At the
early stage of the Kibble-Zurek quench both the vortices and the domain walls
nucleate but eventually the vortices move to the domain walls
and remain trapped there. Thus, amount of unpinned vortices depends on the
vortex--domain wall interaction strength.}

{
    The same reasoning is valid for any multicomponent superconductor which
    supports the formation of the domain walls like $s+id$
    superconductors. In this case a similar magnetic field pattern which
    corresponds to the system of domain walls with the vortices pinned at the
    walls is expected after the Kibble--Zurek quench which complicates
    identification of the chiral $p$-wave superconductivity in the sample.
    However as we noted in the Section III  the magnetic moment of the currents in the films of the non-chiral
    superconductors vanishes so it it possible to distinguish these types of
    pairing performing the measurement of the magnetic moment of the sample
    after the quench.
}

\section{Domain stability}
The vortex--antivortex pairs which appear in the conventional superconductors according to the Kibble--Zurek scenario 
are unstable due to the attraction between the vortices of the opposite
winding numbers. However the impurities in the sample can pin the vortices
thus preventing the relaxation to the homogeneous state. The similar scenario may be relevant for the
domains in the chiral superconductor: the domains can be unstable and shrink
eventually, so the domain picture can be observed only within a finite time after
the quench unless we take pinning into account. Though the total vorticity
of the dominating order parameter component is equal to zero the winding number around
some domains may be nonzero affecting the evolution of the domain. We are
going to discuss the dynamics of the domains
using an
extension of the London theory for the chiral $p$-wave superconductor assuming
$\lambda / \xi \gg 1$.

We restrict ourselves to the 2D case so the domain walls are the contours
which separate the domains of different chirality. The absolute values of the order
parameter components in the bulk of the domain are (1,0) or (0, 1) depending on the
domain type so we can consider the phase of the dominant order parameter
component as a
dynamic variable within the corresponding domains. This gives us the usual
expression for the free energy of the bulk of the domains:
\begin{equation}
    F_{\mathit{bulk}} = \frac{\Phi_0^2}{32 \pi^3 \lambda_{\mathit{eff}}}
    \sum\limits_{\alpha = +,-}
    \int\limits_{\Omega_\alpha} \left(\nabla \theta_\alpha - \frac{2 \pi}{\Phi_0}
    \mathbf A\right)^2\;d^2 r ,
\end{equation}
where $\theta_\pm$ are the phases of the order parameter components and $\Omega_\pm$ are
the areas occupied by the chiral domains.
The domain wall can be viewed as a Josephson junction between the domains
with a certain  equilibrium superconducting phase difference.
The optimal phase difference though depends on the wall orientation as $\theta_+
- \theta_- = 2\theta_n$ for the flat equilibrium walls where $\theta_n$ is the angle between the normal
direction to the wall and the crystal axis\cite{refSigristAgterberg} in
the abscence of tetragonal distortions. Assuming the
curvature of the wall to be much less than $\xi^{-1}$ one can consider the
wall to be almost flat at each point and write the free energy of
the domain wall as follows:
\begin{equation}
    F_{\mathit{wall}} = \frac{\Phi_0^2}{32 \pi^3 \lambda_{\mathit{eff}}} \oint\limits_{\mathit{DW}} \left\{\epsilon + j \cos\left(\theta_+ -
    \theta_+ - 2 \theta_n\right)\right\}\;dl \ ,
\end{equation}
where the integration is taken over all domain walls, $\epsilon$ and $j$
are 
the positive constants which characterize the energy of the domain wall and
the Josephson energy per unit length, respectively.
The
domain walls are energetically unfavorable so the condition $\epsilon > j$ must be
satisfied. The energy of the wall can be obtained
straightforwardly from the Ginzburg--Landau functional by integrating the free
energy density over the short segment across the wall, assuming step--like
form of the absolute values of the order parameter components. Using this approach one
can estimate the parameters $\epsilon \propto \xi^{-1}$ and $j \propto
\zeta \xi^{-1}$, respectively, and find the additional corrections to the wall
energy which come from the phase gradients at the
sides of the wall. These corrections allow to take account of the
supercurrents
flowing along the wall~\cite{refSigristAgterberg} which cannot be described
within the simple Josephson--like model. However in the case of strong type-II
superconductor $\lambda/\xi \gg 1$ and weak interaction between the order
parameter components $\zeta\ll 1$  the Josephson--like
term gives the most significant contribution into the energy of the domain wall.
The free energy of the sample naturally comes as a sum
of bulk and interface terms:
\begin{equation}
    \label{eqLondon}
    F = F_{\mathit{wall}} + F_{\mathit{bulk}} \ .
\end{equation}
This functional yields Laplace equations for the both phases of the order
parameter components with the nonlinear boundary conditions at the domain walls:
\begin{gather}
    \label{eqEulerEqn}
    \nabla^2 \theta_\pm = 0 \ , \\
    \label{eqEulerBnd}
    \left.\frac{\partial \theta_\pm}{\partial n} - j \sin\left(\theta_+ -
    \theta_- - 2 \theta_n\right)\right|_{\mathit{DW}} = 0.
\end{gather}
Here $n$ stands for direction normal to the domain wall from the
``plus'' to the ``minus'' domains.

Using the above model we study the stability of a circular domain of radius
$R$ which
carries no magnetic flux. This requires the abscence of vorticity in the exterior
domain (for certainty we consider $\eta_+$ domain to be an exterior one),
i.e. the phase of the corresponding order parameter component must be a single-valued
function. We neglect the vector potential $\mathbf A$ assuming the sample to
be a thin film and the domain size $R$ to be much less than
the effective penetration length $\lambda_\mathit{eff}$. Due to the nonlinearity of the boundary
conditions~(\ref{eqEulerBnd}) the exact solution of the
equations~(\ref{eqEulerEqn}) appears to be complicated. However in the case of the small
domains and weak interaction between the order parameter components so $j R \ll 1$ one
can linearize the boundary conditions. We suppose that the phases are almost
constant, i.e. 
$|\theta_\pm(\mathbf r) - \Theta_\pm| \ll 1$ for some $\Theta_\pm
= const$. Due to the gauge invariance
an arbitrary constant may be added to both $\Theta_+$ and $\Theta_-$ while
change of the difference $\Theta_+ - \Theta_-$ results in rotation of the
whole domain. Thus without a loss of generality we can assume $\Theta_+ =
\Theta_- = 0$. The phases $\theta_\pm$ must satisfy the Laplace equation
with the following boundaries:
\begin{gather}
    \frac{\partial \theta_+}{\partial r} = j \sin 2 \varphi \ , \\
    \frac{\partial \theta_-}{\partial r} = j \sin 2 \varphi \ .
\end{gather}
Here the angle $\theta_n$ which determines the direction of the normal simply
coincides with the polar angle $\varphi$. One can easily find the solutions
\begin{equation}
    \label{eqLinear}
    \theta_\pm = \mp \frac{j R}{2}
    \left(\frac{r}{R}\right)^{\mp 2} \sin 2 \varphi
\end{equation}
and obtain the free energy of the domain in the lowest order by $R$:
\begin{equation}
    F \approx 2\pi R \epsilon \ .
\end{equation}
The minimum is at $R=0$ which means than the small domains cannot be stable.

The solution~(\ref{eqLinear}) of the equation~(\ref{eqEulerEqn}) for the small domains can be
used as an appropriate ansatz
for the nonlinear problem which appears if the domain is large $jR \gg 1$. We
look for the solution in form of the trial function:
\begin{equation}
    \theta_\pm = \gamma_\pm \left(\frac{r}{R}\right)^{\pm 2} \sin 2
    \varphi\ ,
\end{equation}
where $\gamma_\pm$ are unknown parameters and substitute it into the free
energy~(\ref{eqLondon}):
\begin{equation}
    F = 2\pi\left[\gamma_+^2 + \gamma_-^2 + R \varepsilon + j R J_1 (\gamma_+ - \gamma_-)\right] \ .
\end{equation}
If $j R \gg 1$ the minimum is $\gamma_\pm = \mp x_0 / 2$ where
$x_0 \approx 1.84$
is the position of the first maximum of the Bessel function $J_1(x)$. The
final expression for the free energy is
\begin{equation}
    F \approx 2\pi R \left[\epsilon - j J_1(x_0)\right]
\end{equation}
The dependence also appears to be linear and $dF/dR > 0$ so the domain cannot
be stabilized though the slope of the curve $F(R)$ is reduced compared to the case of the small
domain. 

\begin{figure*}
    \subfigure[]{
        \includegraphics[height=0.48\textwidth]{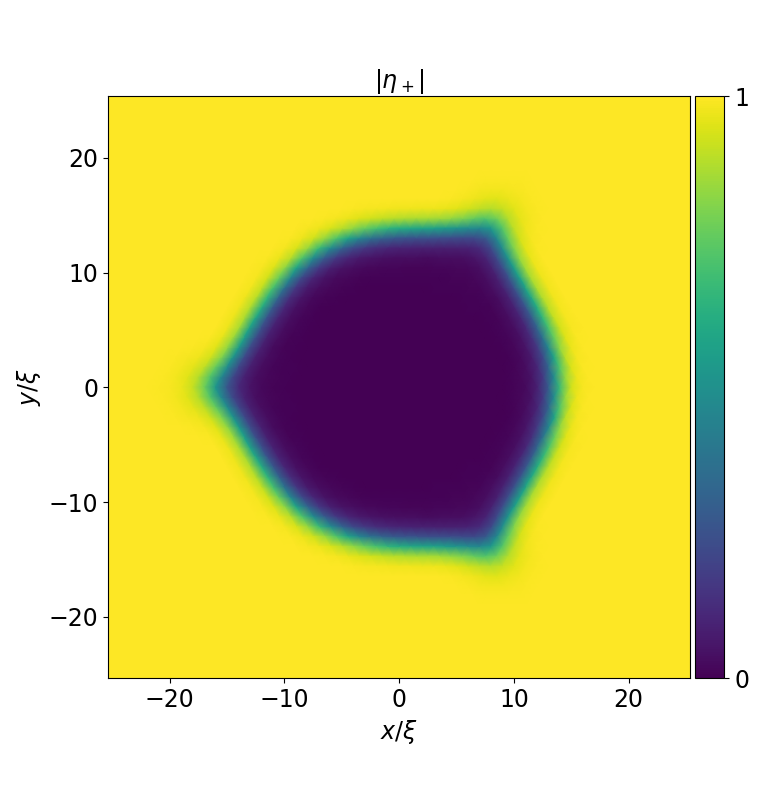}
    }
    \subfigure[]{
        \includegraphics[height=0.48\textwidth]{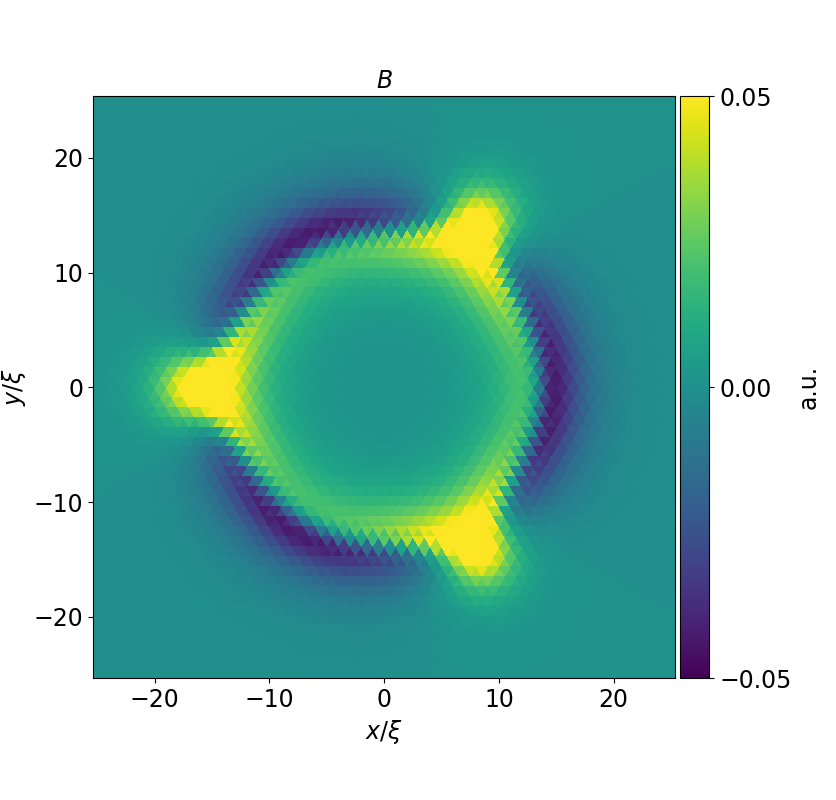}
    }
    \caption{(Color online) The order parameter (a) and the magnetic field (b) of the chiral domain with three vortices pinned by the domain wall.
    The winding number of the outer order parameter component $\eta_+$ is equal to -1.
    The vortices reveal themself as localized peaks of magnetic field. The domain shown on the figure is not stable
    configuration but a snapshot of the domain evolution.}
    \label{figVortices}
\end{figure*}

However the circular domain can be stabilized if it carries two quanta of
magnetic flux. The order parameter of the exterior domain thus has vorticity
equal to $\pm 2$ depending on the domain type. In our case $\theta_- = \pi$
and $\theta_+ = 2 \varphi$. This solution satisfies Laplace equation inside the
domains and the boundary conditions at the domain wall because it minimizes
the Josephson--like energy along the whole wall. The free energy of such
domain is given by the following expression:
\begin{equation}
    F =\frac{\Phi_0^2}{16 \pi^2 \lambda_\mathit{eff}}
            \left[ 4 \ln \frac{\lambda_\mathit{eff}}{R} + (\epsilon - j)
            R\right] \ .
        \end{equation}
The free energy of the exterior domain diverges logarithmically at $r\to
\infty$ so the integral was cutted off at $r = \lambda_\mathit{eff}$. The free
energy has a local minimum at $R_\ast = 4 / (\epsilon - j)$, i.e. the domain
carrying two quanta of the magnetic flux is stable to the radial
perturbations. The numerical simulations
performed within the time dependent
Ginzburg--Landau framework show stability of the two--quanta domains with
respect to the azimuthal perturbations.

The above model may be applied for arbitrary vorticity $n$ of the exterior
order parameter. The presence of nonzero vorticity leads to the logarithmic term
$\propto n^2 \ln (\lambda_\mathit{eff} / R)$ in the free energy expression
which comes from the kinetic energy of the Cooper pairs in the exterior
domain. This term stabilizes the domain at some finite radius.
However in this model all the domains are considered to be circular which is
not true if $n\ne \pm 2$. The Josephson energy is frustrated in this
case and such domains lose circularity due to the azimuthal instability.

This instability reveals itself in appearance of the vortices pinned by the
domain wall.
These vortices represent the short segments of the wall where the phases of the
order parameter components are inconsistent with the Josephson relation. Between these vortices the Josephson--like energy of the
domain wall is minimized. The simulations performed within the time dependent
Ginzburg--Landau model show that these vortices lead to the sharp bending of the
domain wall and loss of the cylindrical symmetry of the domain (see
Fig.~\ref{figVortices}). The azimuthal instability plays crucial role in the
evolution of the domains allowing the domains with $n\ne 0$ to shrink.

\section{Summary}

In this work we have studied the effect of the laser pulse on the film of a
chiral superconductor. Reducing the influence of the laser pulse to the only
effect of the sample heating
 we have found the distribution of the order parameter components and
the magnetic field within the hot spot. We have analyzed the dynamics of the
superconductor after the pulse absorption in the regime of a subsequent
temperature quench. We show that if the initial pulse was strong enough to suppress
superconductivity locally then the chiral domains may grow during the
temperature quench according to the Kibble--Zurek scenario. The magnetic field
created by the currents of the domain walls can be observed experimentally.
The field pattern of the domain walls differs qualitatively from the
field of vortex--antivortex pairs known to appear via Kibble--Zurek mechanism
in the conventional $s$-wave superconductors. Such a behavior is a fingerprint
of the chiral superconductivity and the appropriate experiments may be useful
for it's identification in $\mathrm{Sr_2RuO_4}$.

In order to study the stability of the domains we developed a model which
allows to analyze the samples with the given shape of the domains in London
limit assuming the domain wall to be a Josephson junction with
orientation--dependent Josephson energy. Using this model we studied
stability of the circular domains and show that two--quanta domains are
stable while zero--quanta domains shrink. Simulations within the time dependent
Ginzburg--Landau framework show that the circular domains are unstable with respect to the
azimuthal perturbations if the winding number of the exterior domain differs
from $\pm 2$ due to the Josephson energy frustration {similar to the
frustration in the circular Josephson junctions between the chiral $p$-wave
and the $s$-wave superconductors~\cite{refEtter}.}

We thank I. Shereshevskii, D. Vodolazov, A. Buzdin, Ph. Tamarat and B. Lounis for stimulating discussions. The work was supported by Russian Science Foundation No.
17-12-01383 (ASM), Foundation for the advancement of theoretical physics
``BASIS'' No. 109 (VLV) and Russian
Foundation for Basic Research (VLV).
%

%
%
%
%

%
\end{document}